\documentclass[onecolumn,apj]{emulateapj}
\usepackage{apjfonts}
\usepackage{graphicx}
\usepackage{natbib}

\DeclareGraphicsExtensions{.ps,.png,.eps}

\newcommand{\lsim}{\mathrel{<\kern-1.0em\lower0.9ex\hbox{$\sim$}}}
\newcommand{\gsim}{\mathrel{>\kern-1.0em\lower0.9ex\hbox{$\sim$}}}

\begin{document}

%===============================================================================

%       Title

\title{A Direct Multigrid Poisson Solver for Oct-Tree Adaptive Meshes}

\author{P. M. Ricker\altaffilmark{1,2}
}
\email{pmricker@uiuc.edu
}

\altaffiltext{1}{Department of Astronomy, University of Illinois, Urbana, IL 61801}
\altaffiltext{2}{National Center for Supercomputing Applications, Urbana, IL 61801}

%===============================================================================

%	Abstract

\begin{abstract}
We describe a finite-volume method for solving the Poisson equation on oct-tree
adaptive meshes using direct solvers for individual mesh blocks. The method is
a modified version of the method presented by Huang and Greengard (2000), which
works with finite-difference meshes and does not allow for shared boundaries
between refined patches. Our algorithm is implemented within the FLASH code
framework and makes use of the PARAMESH library, permitting efficient use of
parallel computers. We describe the algorithm and present test results that
demonstrate its accuracy.
\end{abstract}


%===============================================================================

%       Text and acknowledgments

\newpage

%-------------------------------------------------------------------------------
%-------------------------------------------------------------------------------

\section{Introduction}
\label{Sec:intro}

Astrophysical simulations commonly need to solve the Poisson equation,
\begin{equation}
\label{Eqn:Poisson}
\nabla^2 \phi({\bf x}) = 4 \pi G \rho({\bf x})\ ,
\end{equation}
for the gravitational potential $\phi({\bf x})$ given a density distribution
$\rho({\bf x})$. Similar equations also arise in other contexts, such as
incompressible flow problems and divergence-cleaning methods for
magnetohydrodynamics. Self-gravitating problems offer special challenges
because they frequently develop structure spanning large spatial dynamic
ranges. The problem of spatial dynamic range is particularly acute for
grid-based schemes for solving the Euler equations of hydrodynamics.

Within the context of grid-based methods for solving the Poisson equation,
several approaches to the problem of spatial dynamic range have arisen.
The simplest approach is to use Fourier transforms, multigrid methods,
or sparse iterative solvers on uniform Eulerian grids. The maximum dynamic
range is then limited by the available memory. Recently Trac and Pen (2006)
have demonstrated
an out-of-core uniform-grid Poisson solver that exceeds this limit by
making use of disk space; the largest published calculations with this solver
have used $4000^3$ zones. However, storage resource consumption still increases
with the third power of the resolution, putting grids with $10^4$ zones on a
side or larger out of reach for now.

If high resolution is not needed everywhere in the domain, as is frequently
the case in cosmological structure formation simulations, it is also possible
to employ nonuniform Eulerian or Lagrangian grids. Examples include COSMOS
 (Ricker, Dodelson, \& Lamb 2000),
which uses a nonuniform multigrid solver,
and MMH (Pen 1998), which uses a deformable mesh.
These methods work best when the region to be
resolved is known beforehand, although fully Lagrangian codes like Pen's can
follow the development of structures and adjust zone spacing appropriately.
Nonuniform grids, however, introduce complicated position-dependent stencils
and generally cannot be used with fast transform-based solvers. In addition,
coupled numerical hydrodynamics methods generally place constraints on the
allowed mesh anisotropy and nonuniformity, since numerical dissipation increases
with zone spacing.

The greatest spatial dynamic ranges in grid-based astrophysical simulations
have been achieved using adaptive mesh refinement (AMR) techniques. Modern
AMR techniques for solving hyperbolic systems of equations were first developed
by Berger and Oliger (1984) and Berger and Colella (1989). In the Berger and
Colella formulation, AMR involves the construction of a hierarchical set of
mesh ``patches'' with decreasing zone spacing. The coarsest mesh covers the
entire computational domain, while more highly refined meshes cover only a
portion. Generally refined meshes are taken to be nested; that is, each
refined mesh lies completely within its coarser parent mesh. Examples of
astrophysical codes employing patch-based AMR meshes include
the code of Truelove et al.\ (1998),
AMRA (Plewa \& M\"uller 2001),
RIEMANN (Balsara 2001),
Enzo (O'Shea et al.\ 2004),
and CHARM (Miniati \& Colella 2007).
To date self-gravitating AMR calculations have achieved effective
spatial resolutions greater than $10^{15}$.

A considerable simplification of the Berger and Colella method was introduced
by Quirk (1991) and de~Zeeuw and Powell (1993). Known as ``oct-tree'' AMR, this
method requires that each refined patch contain the same number of zones,
that each refinement level have zones a factor of two smaller in each dimension
than the next coarser level, and that each refined patch be no more than one
level removed from its immediate neighbor. Mesh data can then be stored in an
oct-tree data structure, allowing for extremely efficient parallel
implementations, even on high-latency systems (Warren \& Salmon 1993).
Also, because each mesh patch (often
called a ``block'' in this context) contains the same number of zones, it is
possible to achieve high levels of cache re-use when iterating over zones.
Unless each block contains a very small number of zones, this efficiency comes
with the price that refined blocks often must cover more
of the volume than they would in a patch-based method. The primary
astrophysical simulation code employing oct-tree AMR is FLASH (Fryxell
et al.\ 2000), which uses the PARAMESH library (MacNeice et al.\ 2000) to
handle its AMR mesh. (The ART (Kravtsov, Klypin, \& Khoklov 1997),
MLAPM (Knebe, Green, \& Binney 2001), and
RAMSES (Teyssier 2002) codes also use tree
structures to manage AMR meshes, but in these codes the refined blocks
consist of a single zone each, and the base mesh generally contains a large
number of zones, so these codes do not employ oct-trees. A block-based AMR
approach that allows for ``incomplete families'' has also recently been
implemented within the VAC code (van der Holst \& Keppens 2007); the oct-trees
discussed in the current paper require complete ``families'' of child blocks.)
By default PARAMESH
uses blocks containing $8^3$ interior zones as a compromise between adaptive
flexibility and memory efficiency, but any size larger than the differencing
stencil and small enough to fit in the memory attached to a single processor
can be employed.

Poisson solvers on oct-tree meshes generally employ some type of
multigrid or sparse linear solver iteration scheme. Transform methods cannot
be employed directly
because of the varying mesh resolution and non-tensor-product character of
the composite mesh.  For example, Matsumoto and Hanawa (2003) describe a
relaxation-based method for solving the Poisson equation on nested grids.
Within the FLASH framework, we have employed the
Martin and Cartwright (1996) multigrid algorithm for several years. However,
the speed and scalability of this algorithm have been limited because of the
need to apply multiple relaxation iterations on each level, together with the
communication of block boundary data that such iterations require. Because the
cost of this algorithm dominates the cost of most self-gravitating simulations
with FLASH, we are motivated to develop more efficient methods that require
less communication.

In this paper we describe one such method, based on the direct multigrid
algorithm of Huang and Greengard (2000, hereafter HG).
This algorithm improves considerably
upon relaxation-based multigrid solvers by allowing refined patches to be
solved directly using ``black-box'' uniform-grid solvers. Unlike the direct
method described by Couchman (1991), the HG algorithm
properly minimizes the global residual by allowing information to flow back
from fine meshes to coarse meshes. However, it is formulated on a
finite-difference mesh in which refined patches are not permitted to touch.
Here we describe a modified version of the HG algorithm
suitable for finite-volume oct-tree AMR meshes. We have implemented this
algorithm within the FLASH framework, and we present test results that
demonstrate the solver's accuracy. The present paper should be regarded as
a companion to Fryxell et al.\ (2000) and a methodological description for
future FLASH-based simulation papers in the areas of cosmic structure
formation, galaxy cluster physics, star cluster formation, and binary star
evolution, among others.

The paper is organized as follows. In \S~\ref{Sec:algorithm} we give a
precise description of the algorithm. In \S~\ref{Sec:tests} we present the
results of test problems run with the new solver.
We conclude in \S~\ref{Sec:conclusions} with some remarks on performance.
All calculations described
in this paper were performed using version 2.4 of FLASH.

%-------------------------------------------------------------------------------
%-------------------------------------------------------------------------------

\section{The algorithm}
\label{Sec:algorithm}

\subsection{Operator definitions}

Before describing the algorithm, let us first define some terms.
We work with approximations $\tilde\phi({\bf x})$ to the solution to
equation~(\ref{Eqn:Poisson}).
The residual is a loose measure of the error in $\tilde\phi({\bf x})$;
it is given by
\begin{equation}
\label{Eqn:residual}
R({\bf x}) \equiv  4\pi G \rho({\bf x}) - \nabla^2 \tilde\phi({\bf x})\ .
\end{equation}
The first term on the right-hand side is the source
$S({\bf x})$.
Since the Poisson equation is linear, the residual satisfies the equation
\begin{equation}
\nabla^2 C({\bf x}) = R({\bf x})\ ,
\end{equation}
whose solution $C({\bf x})$ is the correction which must be added to
$\tilde\phi({\bf x})$ to yield the correct solution $\phi({\bf x})$.
The source, solution, residual, and correction are all approximated by
zone-averaged values on a hierarchy of mesh blocks.
Where a given mesh block is not a ``leaf node'' --- ie., it is overlain by
another block at a higher level of refinement --- only the residual and
correction are defined (though storage may be allocated for the other
variables as well).
When discussing discretized quantities such as the solution $\tilde\phi$,
we will refer to them in the form $\tilde\phi^{b\ell}_{ijk}$, where $b$
is the block number at the $\ell$th level of refinement ($\ell=1$ being the
coarsest level), and $ijk$ are zone indices within the block $b$.
The notation ${\cal P}(b)$ will refer to the parent (coarser) block
containing block $b$, while ${\cal C}(b,i)$ will refer to
the $i$th child (finer) block associated with $b$ ($i = 1\ldots 2^d$ in
$d$ dimensions). Zone indices are assumed to run between $1\ldots n_x+2n_g$,
$1\ldots n_y+2n_g$,
and $1\dots n_z+2n_g$ in each block, where $n_g$ is the number of ghost zones
(``guard cells'' in PARAMESH parlance) on each boundary.
Between adjacent levels there is a factor
of two in refinement. The generalization to different block/patch sizes and different
refinement factors should be fairly straightforward. A Cartesian mesh
is assumed.

On each block we will also define several operators. 
On level $\ell$, which has zone spacings $\Delta x_\ell$,
$\Delta y_\ell$, and $\Delta z_\ell$ in the $x$-, $y$-, and
$z$-directions,  we define the second-order
difference operator approximating $\nabla^2$ via
\begin{equation}
{\cal D}_\ell\tilde\phi^{b\ell}_{ijk} \equiv
 {1\over\Delta x_\ell^2}\left(\tilde\phi^{b\ell}_{i+1,jk} -
                              2\tilde\phi^{b\ell}_{ijk} +
                              \tilde\phi^{b\ell}_{i-1,jk}\right)+ 
 {1\over\Delta y_\ell^2}\left(\tilde\phi^{b\ell}_{i,j+1,k} -
                              2\tilde\phi^{b\ell}_{ijk} +
                              \tilde\phi^{b\ell}_{i,j-1,k}\right)+ 
 {1\over\Delta z_\ell^2}\left(\tilde\phi^{b\ell}_{ij,k+1} -
                              2\tilde\phi^{b\ell}_{ijk} +
                              \tilde\phi^{b\ell}_{ij,k-1}\right)\ .
\end{equation}
This operator is only applied for $(i,j,k)$ within block interiors.
The difference we use is second-order accurate, but there is nothing about
the HG algorithm that limits us to second order. When
values of $\tilde\phi$ in boundary zones are needed, they are set in one of
three
ways: by restriction from interior zones of a finer neighbor block, by copying
from a neighboring block at the same level of refinement, or by prolongation
from the block's parent. The difference operator is needed for two purposes:
to compute the residual using the finite-volume version of
equation~(\ref{Eqn:residual}), and in the single-block direct solver.

Because we are using finite-volume quantities, we can define the restriction
operator ${\cal R}_\ell$ for block interior zones $(i,j,k)$ exactly as
\begin{equation}
({\cal R}_\ell\tilde\phi)^{{\cal P}(c),\ell}_{ijk} \equiv {1 \over {2^d}}
  \sum_{i'j'k'}
  \tilde\phi^{c,\ell+1}_{i'j'k'}\ ,
\end{equation}
where the indices $(i',j',k')$ refer
to the zones in block $c$ that lie within zone $(i,j,k)$ of block ${\cal P}(c)$.
We apply the restriction operator throughout the interiors of blocks, but its
opposite, the prolongation operator ${\cal I}_\ell$, need only be defined on
the edges of blocks, because it is only used to set boundary values for the
direct single-block Poisson solver:
\begin{equation}
({\cal I}_\ell\tilde\phi)^{c,\ell+1}_{i'j'k'} \equiv \sum_{p,q,r = -2}^2
                 \alpha_{i'j'k'pqr}\tilde\phi^{{\cal P}(c),\ell}_{i+p,j+q,k+r}
\end{equation}
When needed, boundary zone values are set as for the difference operator.
We use conservative quartic interpolation
to set edge values, then solve with homogeneous Dirichlet boundary conditions
after using second-order boundary-value elimination.
(Quadratic interpolants will sometimes work, but experimentation revealed
better convergence with quartic interpolation.)
The coefficients
$\alpha$ determine the interpolation scheme. For the $-x$ face in 3D,
\begin{eqnarray}
\alpha_{1/2,j'k'pqr} &=& \beta_p \gamma_{j'q} \gamma_{k'r} \\
\nonumber
(\beta_p) &=& \left( -{1\over{12}}, {7\over{12}}, {7\over{12}},
                    -{1\over{12}}, 0 \right) \\
\nonumber
(\gamma_{j'q}) &=& \left\{\begin{array}{ll}\displaystyle
                   \left( -{3\over{128}}, {{11}\over{64}},
                                    1, -{{11}\over{64}}, {3\over{128}}\right)
                              & \ \ \ j' {\rm\ odd} \\
                   \displaystyle
                   \left( {3\over{128}}, -{{11}\over{64}},
                                    1, {{11}\over{64}}, -{3\over{128}}\right)
                              & \ \ \ j' {\rm\ even} 
                  \end{array} \right.
\end{eqnarray}
Interpolation coefficients are defined analogously for the other faces.
Note that we use half-integer zone indices to refer to averages over the
faces of a zone; integer zone indices refer to zone averages.

%-------------------------------------------------------------------------------

\subsection{The direct solver and boundary conditions}
\label{Sec:direct solver}

In addition to the above operator definitions, we need a Poisson solver
that can solve problems with given boundary values on uniform Cartesian meshes.
Fortunately, many different algorithms are available to solve this class of
problems, ranging from the easy-to-implement but slow relaxation methods to
fast Fourier transform (FFT) methods and conjugate gradient algorithms.
Indeed, one
of the advantages of the HG algorithm is that it treats the
solution on individual blocks as a `black box.' For our purposes we will use
the $d$-dimensional fast sine transform for homogeneous Dirichlet boundary
conditions with the transform-space Green's function
\begin{equation}
G^\ell_{ijk} =  -16\pi G\left[ {1\over{\Delta x_\ell^2}}\sin^2\left({i\pi\over 2n_x}\right) + 
                       {1\over{\Delta y_\ell^2}}\sin^2\left({j\pi\over 2n_y}\right) +
                       {1\over{\Delta z_\ell^2}}\sin^2\left({k\pi\over 2n_z}\right)\right]^{-1}\ .
\end{equation} 
Note that any second-order solver that can handle homogeneous Dirichlet
boundaries can be modified to work with inhomogeneous (given-value) boundaries
using boundary-value elimination. For example, at the $-x$ boundary, the $x$
part of the difference operator refers to $\tilde\phi_{i-1,j,k}$,
which is unknown for $i=1$. However, if $\tilde\phi(x_{1/2},y_j,z_k)$
is given, then to second order
\begin{equation}
\tilde\phi_{0,j,k} \approx 2\tilde\phi(x_{1/2},y_j,z_k) - \tilde\phi_{1jk}
\end{equation}
(we have dropped the block and level superscripts for clarity).
The second term on the right-hand side is the same for homogeneous boundaries,
so we can recast the problem as a homogeneous one by replacing the source
term as follows:
\begin{equation}
\label{Eqn:bvelim}
\rho_{1jk} \rightarrow \rho_{1jk} - {2\over\Delta x_\ell^2}\phi(x_{1/2},y_j,z_k)\ .
\end{equation}
Corresponding replacements can be made at the other boundaries.

For periodic problems, we need only apply the periodic boundary conditions to
the coarsest (level 1) block. For this purpose we use a real-to-complex FFT
with Green's function
\begin{equation}
G^\ell_{ijk} =  \left\{\begin{array}{ll}\displaystyle
                   {-16\pi G\left[ {1\over{\Delta x_\ell^2}}\sin^2\left({(i-1)\pi\over n_x}\right) + 
    {1\over{\Delta y_\ell^2}}\sin^2\left({(j-1)\pi\over n_y}\right) + 
    {1\over{\Delta z_\ell^2}}\sin^2\left({(k-1)\pi\over n_z}\right)\right]^{-1}}\\
\\
                               \hfill i, j, {\rm\ or\ } k \ne 1 \\
\\
\displaystyle
                   0\hfill i = j = k = 1 
                  \end{array} \right.
\end{equation}
However, this is not all that is required. The Poisson difference operator
with periodic boundary conditions is singular; if the zero-wavenumber
component of the source function is not set exactly to zero, the resulting
solution will not be unique. If the average of the source is zeroed only
for the coarsest block, errors in interpolation to higher levels of refinement
will quickly cause a nonzero DC component to creep into the residual, with
the result that the multigrid V-cycle iteration (described below) will converge
slowly or not at all. Hence when using periodic boundary conditions we explicitly
subtract the average from the residual on all levels, not just the coarsest.
For each level that is not fully refined, we compute the average for that
level by summing over blocks on that level and leaf-node blocks on coarser
levels.

The use of Dirichlet boundaries on individual blocks allows for matching of
the value of the solution across block boundaries at the same level of
refinement, but not its normal derivative. Hence at block corners the
residual converges only at first order. We have tried several
different methods to accelerate the method's convergence at these locations; the
simplest and most effective remedy while preserving second-order differencing
is obtained if we apply two
Gauss-Seidel relaxations to the outer two layers of zones in each block. While this
removes some of the ``directness'' of the method, it is still significantly
faster than using relaxation alone. When periodic boundaries are used, we
subtract the average from the solution following this relaxation step. It is
possible that a higher-order differencing scheme (requiring more boundary
information) could dispense with this additional smoothing.

%-------------------------------------------------------------------------------

\subsection{Multigrid iteration algorithm}

\begin{figure}
\plotone{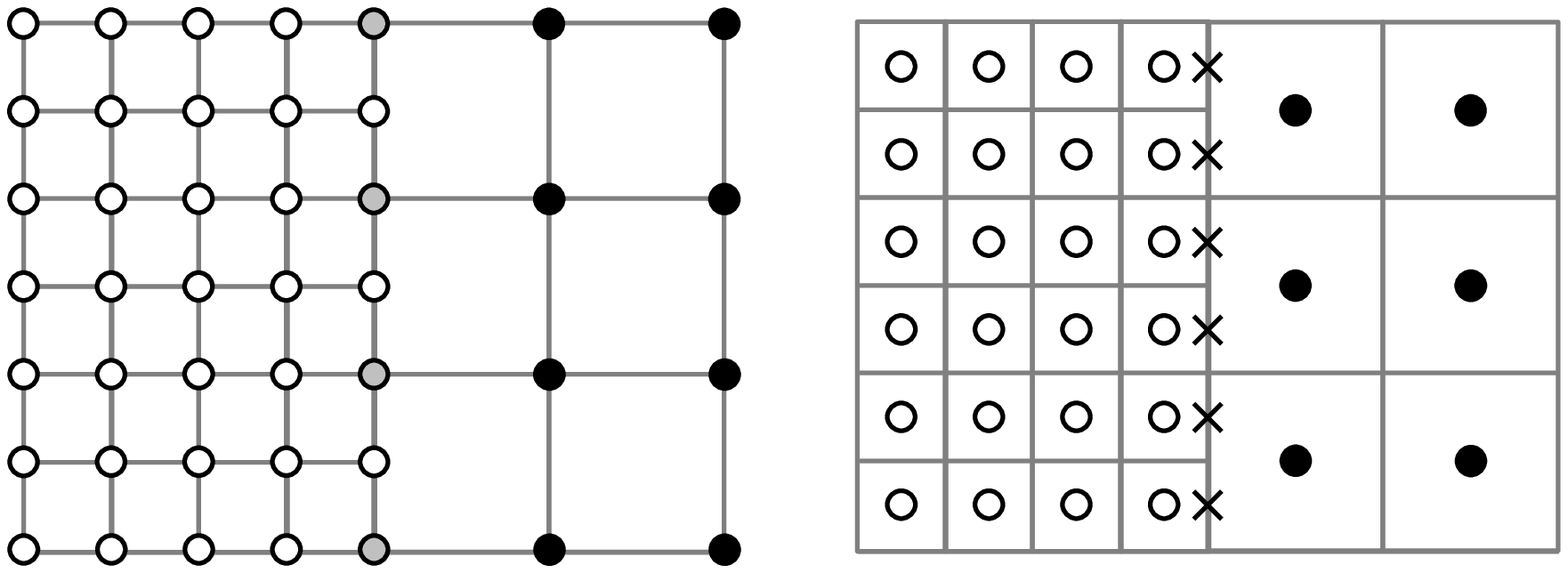}
\caption{\label{Fig:meshes} Mesh structure near a jump in refinement.
Open circles represent fine-mesh points, black circles represent coarse-mesh
points, and gray circles represent points shared by the two meshes.
Left: finite-difference mesh used by HG. Right: finite-volume mesh used in
this paper. Points determined through interpolation are denoted by
$\times$.}
\end{figure}

The two-grid solution procedure for the HG algorithm begins with a coarse-grid
solution $\tilde\phi_{\rm coarse}$
of the discretized Poisson equation. This solution is used to
provide boundary conditions to a finer mesh which may not span the entire
domain (Figure~\ref{Fig:meshes}). Since the original algorithm uses a
finite-difference mesh in which
the coarse and fine meshes share some mesh points, interpolation is needed
only at intermediate points. We will refer to the resulting fine-mesh solution
as $\tilde\phi_{\rm fine}$.

The ``composite'' solution $\tilde\phi_{\rm comp}$ is then taken to be equal to
$\tilde\phi_{\rm coarse}$ on the coarse-grid points and $\tilde\phi_{\rm fine}$
on the fine-grid points. By construction the composite solution is continuous
across the mesh interface, but its normal derivative is not.  This
discontinuity gives rise to a single-layer potential, which when added to the
composite solution yields the corrected solution $\tilde\phi_{\rm corr}$.
HG make use of a lemma
that allows one to compute the single-layer potential simply using the
discontinuity. One first computes a source function $f_{\rm coarse}$ on
the coarse mesh by applying the difference operator on the coarse mesh to 
$\tilde\phi_{\rm coarse} - \tilde\phi_{\rm comp}$ for points on the
mesh interface. Away from the interface, $f_{\rm coarse} = 0$. (Note that no
interpolation is required for this step.) The Poisson
equation is solved on the coarse mesh using $f_{\rm coarse}$ as a source
function, and the result is used to correct $\tilde\phi_{\rm coarse}$. The
corrected coarse-grid solution is then used to provide new boundary
conditions for a corrected fine-grid solution.

The HG algorithm is unsuitable for finite-volume oct-tree meshes for
two reasons. First, as Figure~\ref{Fig:meshes} shows, the coarse and
fine meshes share no discretely sampled points in common. (The finite-volume
mesh stores zone averages, not zone-center values, but to second order
the two are equivalent.) Hence boundary
conditions must be supplied to the fine mesh via interpolation (points
marked with $\times$ in Figure~\ref{Fig:meshes}), and $\tilde\phi_{\rm comp}$
therefore is not guaranteed to be continuous at the boundary. If we wish
to compute the resulting single-layer potential on the coarse mesh, we
must restrict the fine-mesh potential back to the coarse mesh.
The resulting errors in the single-layer potential delay or prevent convergence.
Second, the meshes
used by the HG algorithm do not share boundaries (except possibly an external
boundary). Oct-tree meshes, on the other hand, usually have internal boundaries
that are shared by blocks at the same level of refinement. We wish to apply
the direct solver independently to each block regardless of whether its
boundaries correspond to jumps in refinement.

We can address both issues by realizing that the single-layer potential
source $f_{\rm coarse}$ can be replaced with a residual if the direct solver
reduces the residual to machine zero away from jumps in refinement. For
example, consider the mesh depicted at the left of Figure~\ref{Fig:meshes},
taking $\Delta x$ to be the fine-mesh spacing and indexing points on both
meshes using the same system. The points along
the mesh interface correspond to $i=0$.
Hence the first column of coarse-grid points to the
right of the interface is at $i=2$, while the first column of fine-grid
points to the left of the interface is at $i=-1$. For this situation,
$f_{\rm coarse}$ is given by
\begin{equation}
f_{{\rm coarse},ij} =  \left\{\begin{array}{ll}\displaystyle
                   {{\tilde\phi_{{\rm coarse},-2,j} - \tilde\phi_{{\rm fine},-2,j}}
                      \over {4\Delta x^2}}
                              & i = 0 \\
\\
\displaystyle
                   0
                              & i \ne 0
                  \end{array} \right.
\end{equation}
The residual in $\tilde\phi_{\rm comp}$, on the other hand, is for points on
the boundary,
\begin{equation}
R_{{\rm comp},0,j} = 4\pi G\rho_{{\rm coarse},0,j}   -
     {{\tilde\phi_{{\rm fine},-2,j} - 2\tilde\phi_{{\rm coarse},0,j} +
       \tilde\phi_{{\rm coarse},2,j}} \over {4\Delta x^2}}  - 
 {{\tilde\phi_{{\rm coarse},0,j-2} - 2\tilde\phi_{{\rm coarse},0,j} +
       \tilde\phi_{{\rm coarse},0,j+2}} \over {4\Delta y^2}}
     \ ,
\end{equation}
whereas for $i > 0$ it is effectively zero. 
(It is not necessarily zero for $i < 0$, since the coarse-grid difference
operator would be applied to $\tilde\phi_{\rm fine}$ there.)
The coarse-grid solution was
obtained using the direct solver, so the coarse-grid residual on the interface
is also effectively zero:
\begin{eqnarray}
R_{{\rm coarse},0,j} &=& 4\pi G\rho_{{\rm coarse},0,j}   -
     {{\tilde\phi_{{\rm coarse},-2,j} - 2\tilde\phi_{{\rm coarse},0,j} +
       \tilde\phi_{{\rm coarse},2,j}} \over {4\Delta x^2}}   -
  {{\tilde\phi_{{\rm coarse},0,j-2} - 2\tilde\phi_{{\rm coarse},0,j} +
       \tilde\phi_{{\rm coarse},0,j+2}} \over {4\Delta y^2}}\\
\nonumber
 & =& 0
\ .
\end{eqnarray}
Combining these equations to eliminate $4\pi G\rho_{{\rm coarse},0,j}$ yields
\begin{equation}
R_{{\rm comp},0,j} = {{\tilde\phi_{{\rm coarse},-2,j} - \tilde\phi_{{\rm fine},-2,j}}
                      \over {4\Delta x^2}}\ ,
\end{equation}
which is the same as $f_{{\rm coarse},0,j}$. If $R_{{\rm comp},ij}$ is forced
to be zero also for $i < 0$ (e.g.\ by computing the residual on the fine mesh
and restricting it), the correction generated by solving the Poisson
equation with $R_{{\rm comp},ij}$ as a source function should therefore be
equal to the single-layer potential.

We can use this insight to create a finite-volume version of the HG
algorithm that uses the residual to construct the correction. Referring to
the right part of Figure~\ref{Fig:meshes}, we begin by solving for
$\tilde\phi_{\rm coarse}$ on the coarse mesh, subject to the external
boundary conditions, and use conservative quartic interpolation on
$\tilde\phi_{\rm coarse}$ to set boundary values at the mesh interface.
We solve for $\tilde\phi_{\rm fine}$ on the fine mesh with these boundary
conditions and compute the residual on the fine mesh.

Computing the residual
in the fine-mesh zones adjacent to the interface requires boundary
values interpolated from the coarse mesh; we again use quartic interpolation,
but instead of interpolating interface values, we interpolate average values
over the quarter of each coarse-mesh zone adjacent to each fine-mesh zone.
This allows us to use the same centered difference operator as implied by
the Green's function. It also allows us
to treat boundaries involving jumps in refinement in the same way as 
boundaries separating blocks at the same refinement level.

We next restrict the fine-mesh residual to the part of the coarse mesh
that underlies it and compute the coarse-mesh residual throughout the
remainder of the coarse mesh. We should now have a very small residual
everywhere except for the coarse-mesh zones just to the left of the mesh
interface (ie.\ underlying the fine mesh). We solve the correction
equation on the coarse mesh using this residual as a source function,
subject to homogeneous Dirichlet boundary conditions. We use the solution
to correct
$\tilde\phi_{\rm coarse}$, interpolate the coarse-mesh correction to the mesh
interface, and use these boundary conditions to solve the correction equation
on the fine mesh. Applying the fine-mesh correction to $\tilde\phi_{\rm fine}$
completes the two-grid iteration. We recompute the residual on both
levels and use its
norm to determine whether to apply another correction.

We can construct an algorithm for arbitrary numbers of refinement levels by
recursively extending our two-grid iteration.
In terms of the operators defined previously, here are the steps in our
modified HG algorithm:

\begin{enumerate}
\item Begin by restricting the source function $4\pi G\rho$
      on all levels so that it is
      defined on every block.  If using periodic boundary conditions,
      subtract the global average.
\item {\it Interpolation step:} For each level $\ell$ from 1 to the maximum
      level $\ell_{\rm max}$,
      \begin{enumerate}
      \item If $\ell = 1$ and using external Dirichlet boundaries, set external
            boundary (face) values of $\tilde\phi^{b\ell}_{ijk}$
            for all blocks $b$ on level $\ell$ to 0.
      \item Solve ${\cal D}_\ell \tilde\phi^{b\ell}_{ijk} =
            4\pi G\rho^{b\ell}_{ijk}$ for all blocks $b$ on level $\ell$.
            (The edge relaxation mentioned in \S~\ref{Sec:direct solver} is
            applied immediately after the direct solver here.)
      \item Compute the residual $R^{b\ell}_{ijk} = 4\pi G\rho^{b\ell}_{ijk} -
            {\cal D}_\ell \tilde\phi^{b\ell}_{ijk}$ for all blocks $b$ on
            level $\ell$. In setting boundary values for this step, ignore
            levels $>\ell$, copy data from neighboring blocks at level $\ell$,
            and interpolate from neighboring blocks at levels $<\ell$.
      \item For each block $b$ on level $\ell$ that has children, interpolate
            face boundary values of $\tilde\phi^{b\ell}_{ijk}$ for each child.
      \end{enumerate}
\item {\it Residual propagation step:} 
      Restrict the residual on all levels so that $R^{b\ell}_{ijk}$ contains
      either a ``coarse grid'' residual if $b$ is a leaf-node block or a
      restricted ``fine grid'' residual if $b$ is not.
\item {\it Correction step:} Compute the discrete L2 norm of the residual over
      all leaf-node blocks and divide it by the discrete L2 norm of the source
      over the same blocks. If the result is greater than a preset threshold
      value, proceed with a correction step: for each level $\ell$ from 1 to
      $\ell_{\rm max}$,
      \begin{enumerate}
      \item If $\ell = 1$, set external boundary (face) values of
            $C^{b\ell}_{ijk}$ for all blocks $b$ on level $\ell$ to 0.
      \item Solve ${\cal D}_\ell C^{b\ell}_{ijk} =
            R^{b\ell}_{ijk}$ for all blocks $b$ on level $\ell$.
            (The edge relaxation mentioned in \S~\ref{Sec:direct solver} is
            applied immediately after the direct solver here.)
      \item Overwrite $R^{b\ell}_{ijk}$ with the new residual
            $R^{b\ell}_{ijk} - {\cal D}_\ell C^{b\ell}_{ijk}$ for all blocks
            $b$ on level $\ell$. In setting boundary values for this step, ignore
            levels $>\ell$, copy data from neighboring blocks at level $\ell$,
            and interpolate from neighboring blocks at levels $<\ell$.
      \item Correct the solution on all leaf-node blocks $b$ on level $\ell$:
            $\tilde\phi^{b\ell}_{ijk} \rightarrow \tilde\phi^{b\ell}_{ijk} +
            C^{b\ell}_{ijk}$.
      \item For each block $b$ on level $\ell$ that has children, interpolate
            face boundary values of $C^{b\ell}_{ijk}$ for each child.
      \end{enumerate}
\item If a correction step was performed, return to the residual propagation
      step.
\end{enumerate}

\noindent
The above procedure requires storage for $\tilde\phi$, $C$, $R$, and $\rho$ on
each block, for a total storage requirement of $4 n_x n_y n_z$ floating-point
values per block (not including boundary zones). Although it is expressed as
a serial algorithm, a parallel version can be constructed straightforwardly (as
in the FLASH implementation) by using a space-filling curve to allocate blocks
to processors. Parallel communication is required when interpolating,
restricting, or copying boundary values between blocks that are stored on
different processors.  Computation of L2 norms and averages also
requires a global reduction operation.

It should be emphasized that our algorithm differs in a crucial way from the HG
algorithm in its treatment of single-layer potential propagation from fine to
coarse meshes. Rather than explicitly propagating single-layer potentials from
level to level and computing new single-layer sources as we go (as in HG's Step~3),
we propagate residuals from the previous interpolation or correction step and
rely on nested iteration to communicate the fine-mesh corrections throughout the
mesh. Although this method converges more slowly than the HG algorithm, it
is much easier to code and parallelize for an oct-tree mesh.

%-------------------------------------------------------------------------------
%-------------------------------------------------------------------------------

\section{Test results}
\label{Sec:tests}

In this section we describe the results of two astrophysically motivated
test problems solved using the modified HG algorithm as implemented within
FLASH version 2.4:  the potential of a homogeneous oblate spheroid, with isolated
boundary conditions, and a snapshot from
a simulation of galaxy cluster formation within a $\Lambda$CDM universe,
using periodic boundary conditions.

%-------------------------------------------------------------------------------

\subsection{The potential of a homogeneous spheroid}
\label{Sec:ellipsoid potential test}

The potential of a homogeneous ellipsoid was first derived independently
by Gauss (1813) and Rodrigues (1815). It is used in the construction of
equilibrium solutions for rotating fluid bodies, such as the sequences of
Maclaurin spheroids and Jacobi and Dedekind ellipsoids.

Consider a homogeneous ellipsoid of density $\rho$ centered on the
origin with principal axes
$a_1$, $a_2$, and $a_3$ aligned with the $x$-, $y$-, and $z$-axes,
respectively. The gravitational potential at a point
${\bf x} = (x,y,z)$ interior to the ellipsoid is (Chandrasekhar 1969)
\begin{equation}
\label{Eqn:ellipsoid potential start}
\phi({\bf x}) = -\pi G\rho\left[A_1(a_1^2-x^2) + A_2(a_2^2-y^2) +
                A_3(a_3^2-z^2)\right]\ ,
\end{equation}
where
\begin{eqnarray}
\nonumber
A_i &\equiv& a_1a_2a_3\int_0^\infty {du\over\Delta(a_i^2+u)}\ \ (i=1,2,3) \\
\Delta &\equiv& \left[(a_1^2+u)(a_2^2+u)(a_3^2+u)\right]^{1/2}\ .
\end{eqnarray}
 In general the expressions for $A_i$
can be written in terms of incomplete elliptic integrals, but for oblate
spheroids, $a_1 = a_2 > a_3$, and the expressions simplify
to
\begin{eqnarray}
\nonumber
A_1 = A_2 & = & {\sqrt{1-e^2}\over e^3}\sin^{-1}e - {1-e^2\over e^2}\\
\label{Eqn:oblate A}
A_3 & = & {2\over e^2} - {2\sqrt{1-e^2}\over e^3}\sin^{-1}e\ ,
\end{eqnarray}
where
\begin{equation}
e = \sqrt{1 - \left({a_3\over a_1}\right)^2}
\end{equation}
is the ellipticity of the spheroid. If the point ${\bf x}$ is external to
the ellipsoid, the potential is given by
\begin{equation}
\phi({\bf x}) = \pi G\rho a_1a_2a_3\int_\lambda^\infty{du\over\Delta}
                \left(1-{x^2\over a_1^2+u}-{y^2\over a_2^2+u}-{z^2\over a_3^2+u}
                \right)\ ,
\end{equation}
where $\lambda$ is the positive root of the equation
\begin{equation}
{x^2\over a_1^2+\lambda}+{y^2\over a_2^2+\lambda}+{z^2\over a_3^2+\lambda}=1\ .
\end{equation}
For oblate spheroids this simplifies to
\begin{equation}
\phi({\bf x}) = -{2 a_3\over e^2}\pi G\rho\left[a_1e\tan^{-1}h -
                {1\over2}\left(\left(x^2+y^2\right)\left(\tan^{-1}h -
                   {h\over 1+h^2}\right) + 2z^2\left(h - \tan^{-1}h\right)
                   \right)\right]\ ,
\end{equation}
where
\begin{equation}
\label{Eqn:ellipsoid potential end}
h\equiv {a_1e\over\sqrt{a_3^2 + \lambda}}
\end{equation}
(Broucke \& Scheeres 1994).

\begin{figure}
\plotone{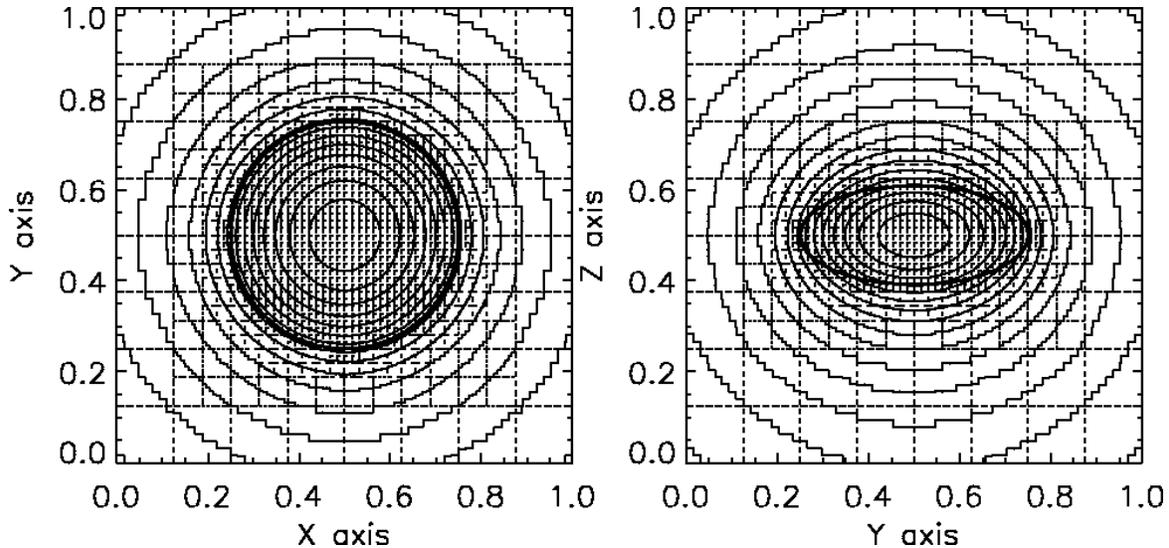}
\caption{\label{Fig:spheroid potential} 
Potential of a Maclaurin spheroid with eccentricity 0.9 as computed using
an adaptively refined mesh with 6 levels of refinement. The left image
shows potential contours (solid), AMR block outlines (dotted), and the
surface of the ellipsoid (thick solid) in the $xy$ plane passing through the
center of the ellipsoid. The right image shows the same quantities for
the $yz$ plane. Note that potential contours cross the ellipsoid surface
in the $yz$ plane because the ellipsoid is an equilibrium figure only when
rotating.
}
\end{figure}

Using our modified HG solver within FLASH, we computed potentials
for oblate spheroids of fixed density ($\rho = 1$), fixed semimajor axis
($a_1=0.25$), and varying eccentricity ($e = 10^{-6}$, 0.5, 0.96).
The spheroids were centered in a box with unit dimensions.
For this test we used the known analytical solution to set external
boundary conditions in order to study the convergence properties of the
multigrid solver by itself.
(For general isolated problems we use a variant of James' (1977) image-mass method
to implement the boundary conditions. This involves two calls to the
Poisson solver, plus an external boundary value computation for the image
mass distribution. This boundary value computation is carried out in FLASH
using a separate multipole Poisson solver.)
We used maximum refinement levels between 1 and 6, with blocks containing
$8^3$ zones, giving $8^3$ -- $256^3$ effective resolution.
An example of the potential for $e = 0.9$, computed with maximum
refinement level 6, is shown in Figure~\ref{Fig:spheroid potential}.
For partially refined calculations, blocks that contain any points
within the spheroid are marked for refinement. This type of refinement
is frequently beneficial when solving azimuthally symmetric problems
on Cartesian meshes.

Since this problem has an analytic solution, we can examine
the absolute convergence properties of the solver.
Here we use the term ``convergence'' in two
ways. The {\it iteration convergence rate} is the rate at which
multigrid V-cycles converge to a solution and is given by
${\cal E}[R_{n}]/{\cal E}[R_{n-1}]$, where $R_n$ is the residual after
$n$ V-cycles.
Following Briggs et al.\ (2000), we define the L2 norm of a
quantity $u^b_{ijk}$ defined on zones $ijk$ of block $b$ to be
\begin{equation}
\label{Eqn:L2 norm}
{\cal E}[u] \equiv \left[{1\over V}\sum_{ijkb} ({u^b_{ijk}})^2\Delta V^b_{ijk}
   \right]^{1/2}\ ,
\end{equation}
where $V$ is the volume of the domain, $\Delta V^b_{ijk}$ is the volume of
the $ijk$th zone of block $b$, and the sum is taken over all leaf-node
blocks and their interior zones.
For this test, unless otherwise specified, we allowed the solver to
iterate until the residual norm dropped below $10^{-15}$ of the source
norm.
The {\it mesh convergence rate}, on the other hand, is the rate at which
the solution converges to the correct solution on meshes of decreasing
minimum zone size. We compute the mesh convergence
rate by injecting adaptive-mesh solutions at different maximum
refinement levels onto a $512^3$ uniform mesh, then comparing them to the
analytical solution computed on the same mesh.
The error should scale as
$\Delta x_{\rm min}^p$, where $\Delta x_{\rm min}$ is the smallest
zone spacing on each adaptive mesh and $p$ is the mesh convergence rate.
For uniform meshes this error should be simply
the truncation error of the differencing scheme used to solve the
Poisson equation. Ideally our adaptive mesh scheme should not degrade
this performance significantly.

\begin{figure}
\epsscale{0.75}
\plotone{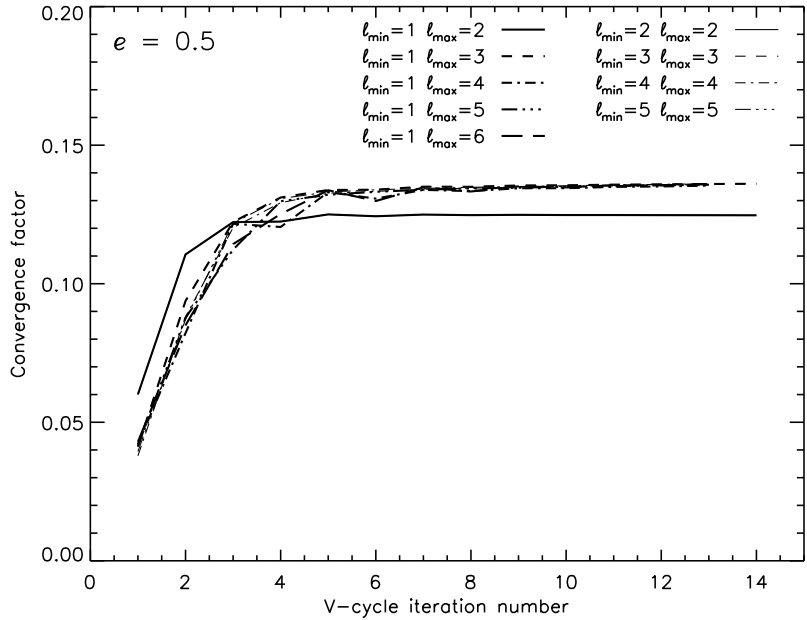}
\caption{\label{Fig:spheroid conv} 
Convergence factor as a function of iteration count for the $e = 0.5$
Maclaurin spheroid potential on meshes with different minimum and maximum
levels of refinement $\ell_{\rm min}$ and $\ell_{\rm max}$.
}
\end{figure}

Figure~\ref{Fig:spheroid conv} shows the iteration convergence
for the $e = 0.5$ case. (The other cases show nearly identical behavior.)
The residual is
reduced by about a factor of twenty during the first V-cycle; thereafter
the convergence factor increases before flattening to
an asymptotic value of about 0.135. The overall convergence behavior does not
vary significantly with maximum refinement level; each case reaches the
single-precision convergence threshold ($10^{-6}$) within three iterations and the
$10^{-10}$ level within seven iterations. This behavior --- initially
rapid reduction of residuals followed by approach to a resolution-independent
convergence factor --- is typical for multigrid algorithms (Briggs et al.\ 2000);
indeed, it is what permits multigrid algorithms to achieve optimal problem-size
scaling for elliptic problems.

\begin{figure}
\epsscale{0.75}
\plotone{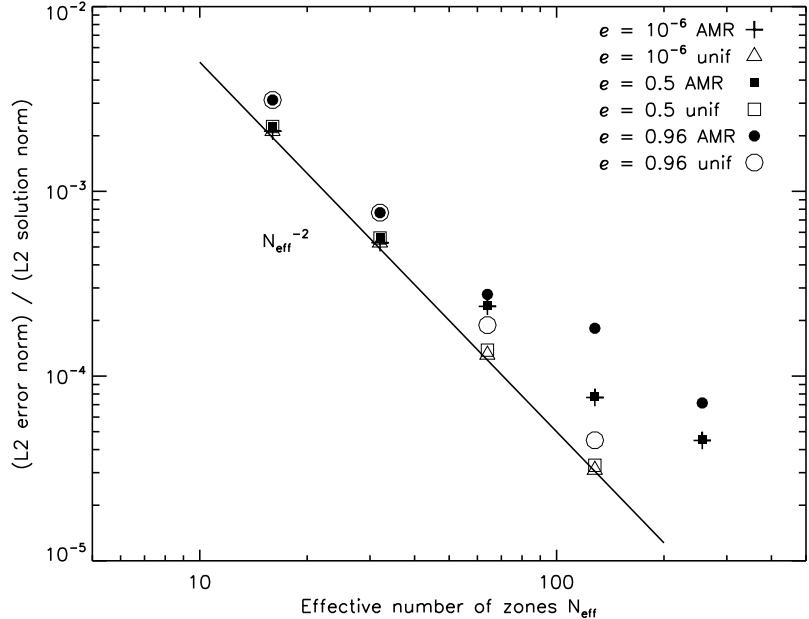}
\caption{\label{Fig:spheroid error}
Error in numerical solutions for the potentials of Maclaurin spheroids
of varying eccentricity, as functions of the effective resolution $N_{\rm eff}$.
We plot the norm of the error divided by the norm of the analytical solution.
}
\end{figure}

Although iteration convergence is necessary for a usable algorithm, it
does not by itself guarantee a correct solution.
In Figure~\ref{Fig:spheroid error} we show the mesh convergence of
our numerical solution relative to the analytical solution
(equations (\ref{Eqn:ellipsoid potential start}) --
(\ref{Eqn:ellipsoid potential end}))
for both uniformly refined and partially refined meshes.
As expected, the convergence of the uniformly refined meshes reflects
the ${\cal O}(\Delta x^2)$ truncation error of our differencing scheme.
For partially refined meshes, the convergence rate
depends on the fraction of the volume occupied by zones
adjacent to jumps in refinement. For $e = 10^{-6}$ and $e = 0.5$ at
5 levels of refinement, this fraction is about 4.7\%, and the error
converges as $\Delta x^{1.2}$ when compared with the error at 4 levels
of refinement. For $e = 0.96$ about 13\% of the volume
is adjacent to refinement jumps, and the error converges as $\Delta x^{1.0}$.
For six levels of refinement the fractions change to 8.8\% and 6.4\%,
respectively, and this change is reflected in a decrease in the
convergence rate for $e = 10^{-6}$ and $e = 0.5$ and an increase for
$e = 0.96$.

%-------------------------------------------------------------------------------

\subsection{Cosmological simulation snapshot}

The cosmological simulation test uses a snapshot from the
``small box'' galaxy cluster
simulation described by Heitmann et al.\ (2005). This simulation
followed the evolution of a $(64h^{-1}\ {\rm Mpc})^3$ volume in a spatially
flat $\Lambda$CDM cosmology with matter density parameter $\Omega_{m0} = 0.314$,
power spectrum normalization $\sigma_8 = 0.84$,
and Hubble parameter $h=0.71$. To examine the self-convergence of the
solver, we ran the test on several different partially- and fully-refined
meshes and compared them with a $1024^3$ uniform-mesh solution obtained
using the same solver.. The dataset consists of $256^3$ particles,
each with mass $1.918\times10^9\ M_\odot$, at redshift zero. Periodic
boundary conditions were used.

We computed potentials using FLASH with both partially refined and fully
refined (uniform) meshes.
To determine whether to refine each block in the partially refined cases,
we computed the number of
particles in each zone and took the maximum of the values within the block.
Blocks with a maximum number of particles per zone greater than 200 were allowed
to refine. We repeated the refinement step until no further refinements
occurred. This refinement criterion corresponds to a varying-interval logarithmic
density threshold. While other density-based criteria are possible, it is
beyond the scope of this paper to choose an optimal criterion; we will present
such a comparison in a future paper.

\begin{figure}
\epsscale{1.1}
\plotone{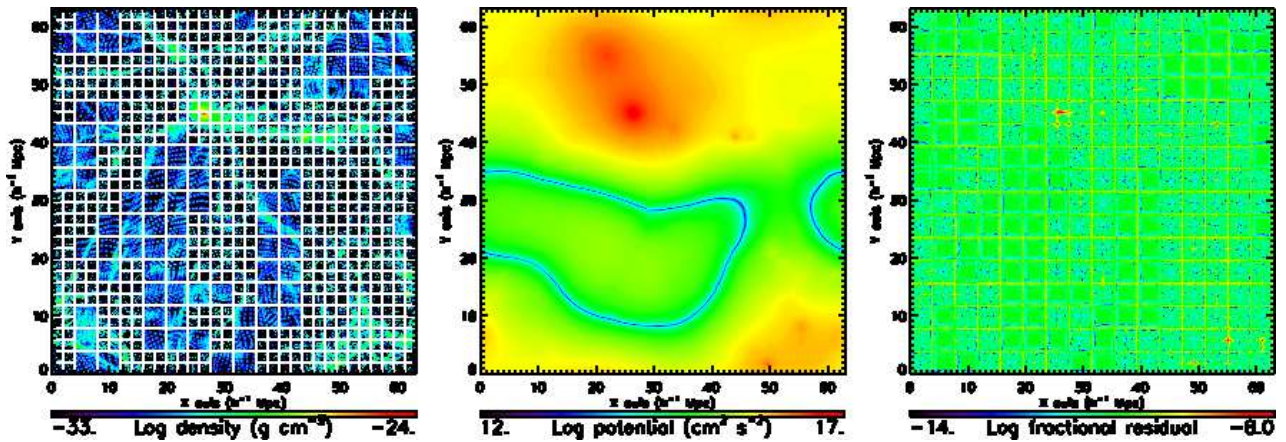}
\caption{\label{Fig:lcdm slice}
The $\Lambda$CDM snapshot test.
Left, distribution of dark matter particles within a slice $125 h^{-1}\ {\rm kpc}$
thick containing the most massive halo at zero redshift in the ``small box'' test of
Heitmann et al.\ (2005). White lines indicate block boundaries; seven levels
of refinement were used.
Middle, potential computed using the modified HG
algorithm in the same slice.  Right, residual as a fraction of the $G$ times the
maximum
density ($1.8\times10^{-25}{\rm\ g\ cm}^{-3}$) in the same slice.
}
\end{figure}

Figure~\ref{Fig:lcdm slice} shows the distribution of dark matter particles
within a slice $125h^{-1}\ {\rm kpc}$ thick passing through the most massive
halo in this dataset. Also shown are the potential and the residual in the
same slice as computed using the modified HG
algorithm in FLASH with seven levels of refinement. Using blocks containing
$16^3$ zones, this corresponds to an effective resolution of $1024^3$.
The potential plot shows the expected correspondence of local potential wells
with massive clusters. The residual plot shows that most of the residual is
contained in the zones adjacent to block boundaries, with local peaks at
the corners of maximally refined blocks in high-density regions. Throughout
most of the volume, however, the residual is less than $10^{-10}$ of the
maximum density, despite the fact that the termination criterion specified
that the residual norm drop only below $10^{-6}$ of the source norm. The residual
can be reduced to double-precision round-off level everywhere through
additional correction
cycles, but this level is adequate for most simulation purposes.

\begin{figure}
\epsscale{0.75}
\plotone{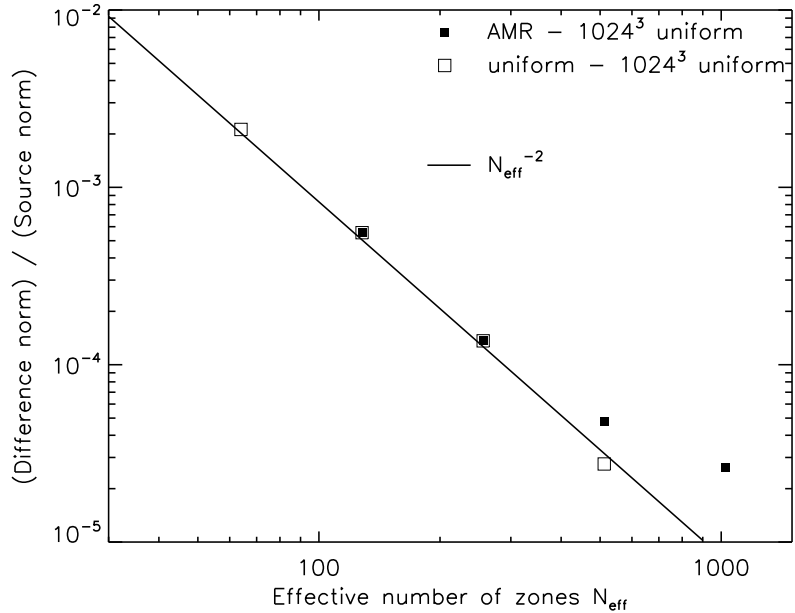}
\caption{\label{Fig:lcdms error}
Mesh convergence in the $\Lambda$CDM snapshot test. Plotted is the
L2 norm of the difference between the potential computed using the modified HG
algorithm and the uniform-grid $1024^3$
modified HG result (squares). Results from partially refined AMR meshes
(filled symbols) and uniformly refined meshes (open symbols) are shown.
}
\end{figure}

In Figure~\ref{Fig:lcdms error} we examine the mesh convergence of
the modified HG solver. The FLASH potential from each run
was injected onto a uniform $1024^3$ mesh for this comparison.
The self-difference norm shows the expected second-order convergence in both
the partially refined and uniformly refined cases up to $512^3$ effective
resolution. At $1024^3$ the use of
density-based refinement degrades the convergence rate for partially refined
meshes to first order. Iteration convergence results for this periodic
problem are nearly identical to those for the isolated Maclaurin spheroid
problem discussed in the previous section; the $10^{-6}$ level is always
reached within five iterations.

%-------------------------------------------------------------------------------
%-------------------------------------------------------------------------------

\section{Conclusions}
\label{Sec:conclusions}

We have detailed the modifications needed in order to use the Huang \& Greengard (2000)
algorithm for Poisson's equation on oct-tree AMR meshes. This method allows us
to use a local direct Poisson solver with Dirichlet boundary conditions on each
block, yet it correctly minimizes the residual across the composite AMR mesh.
The HG algorithm must be modified to take into account the fact that mesh quantities
represent zone-averaged values and the fact that block boundaries can coincide on
an oct-tree mesh. Because adjacent coarse and fine blocks do not share points in
common for oct-tree meshes, additional interpolation (as compared to HG) is needed to set
boundary values. The resulting errors at block corners reduce convergence to
first order, but a fixed small number of boundary-zone relaxation
steps restores the desired second-order convergence rate for block corners in
uniformly refined regions. A higher-order scheme
may be able to eliminate the need for such relaxation.
Our test results show that, while jumps in refinement degrade convergence somewhat
in comparison with solving on a uniform mesh, the effect is manageable because
oct-tree meshes are usually fully refined up to some level. Thus first-order
convergence takes over only once the error has been significantly reduced at second
order on fully refined levels. 

The parallel scaling of this solver, as implemented using the Message Passing
Interface (MPI) in the FLASH code, is comparable to or slightly better than that
of the relaxation
multigrid solver distributed with FLASH 2.$x$. With constant total work and increasing
processor count, parallel efficiency is close to 100\% up to 8 -- 16 times the
smallest number of processors on which a run can fit. When gasdynamics is
included, the Poisson solver requires $\sim$50\% of the execution time; for particle-only
simulations the amount is $\sim $70 -- 80\%. In comparison with the older solver,
a factor of 2 -- 5 improvement in performance is often seen. The solver described
here will be made available to the public as part of FLASH~3.0.\footnote{FLASH is
freely available at http://flash.uchicago.edu/.}

The primary scaling bottleneck for this and other multigrid algorithms is the
fact that on the coarsest level there are too few zones to distribute among all
of the processors. Since each V-cycle descends to the coarsest level, processors
controlling blocks on finer levels must wait until the coarsest level is finished
before proceeding. To counteract this work starvation,
we are investigating the use of a uniform-grid
parallel FFT solver (T.~Theuns, private communication) to handle the coarsest level
in a distributed fashion. Additional strategies may be necessary to make optimal
use of the petaflop computing resources now beginning to become available.

%-------------------------------------------------------------------------------
%-------------------------------------------------------------------------------

%       Acknowledgments

\acknowledgments
The author would like to thank T.~Plewa for calling his attention to the
Huang and Greengard paper and the anonymous referee for helpful comments.
Portions of this work were completed at the Aspen Center for Physics.

This work was supported by the University of Illinois at Urbana-Champaign and
the National Center for Supercomputing Applications, which also provided
supercomputing resources. Additional support was
provided under a Presidential Early Career Award for Scientists and Engineers
by Lawrence Livermore National Laboratory contract LLL B532720.
Development of FLASH was supported by DOE under contract B341495 to the
ASC Center for Astrophysical Thermonuclear Flashes at the University
of Chicago.

%===============================================================================

%       References

\newcommand{\and}{\ \&}

%===============================================================================

\end{document}